\documentclass[aps, prb, twocolumn,showpacs,superscriptaddress]{revtex4-1}

\usepackage{amsmath,amssymb}
\usepackage{graphicx}% Include figure files
\usepackage{dcolumn}% Align table columns on decimal point
\usepackage{bm}% bold math

\begin{document}

\title{Extended dual description of Mott transition beyond two-dimensional space}

\author{Yin Zhong}
\affiliation{Center for Interdisciplinary Studies $\&$ Key Laboratory for
Magnetism and Magnetic Materials of the MoE, Lanzhou University, Lanzhou 730000, China}
\author{Ke Liu}
\affiliation{Institute of Theoretical Physics, Lanzhou University, Lanzhou 730000, China}
\author{Hong-Gang Luo}
\affiliation{Center for Interdisciplinary Studies $\&$ Key Laboratory for
Magnetism and Magnetic Materials of the MoE, Lanzhou University, Lanzhou 730000, China}
\affiliation{Beijing Computational Science Research Center, Beijing 100084, China}

\date{\today}

\begin{abstract}
Motivated by recent work of Mross and Senthil [Phys. Rev. B \textbf{84}, 165126 (2011)] which provides a dual description for Mott transition from Fermi liquid to quantum spin liquid in two space dimensions, we extend their approach to higher dimensional cases, and we provide explicit formalism in three space dimensions. Instead of the vortices driving conventional Fermi liquid into quantum spin liquid states in 2D, it is the vortex lines to lead to the instability of Fermi liquid in 3D. The extended formalism can result in rich consequences when the vortex lines condense in different degrees of freedom. For example, when the vortex lines condense in charge phase degrees of freedom, the resulting effective fermionic action is found to be equivalent to that obtained by well-studied slave-particle approaches for Hubbard and/or Anderson lattice models, which confirm the validity of the extended dual formalism in 3D. When the vortex lines condense in spin phase degrees of freedom, a doublon metal with a spin gap and an instability to the unconventional superconducting pairing can be obtained. In addition, when the vortex lines condense in both phase degrees, an exotic doubled $U(1)$ gauge theory occurs which describes a separation of spin-opposite fermionic excitations. It is noted that the first two features have been discussed in a similar way in 2D, the last one has not been reported in the previous works. The present work is expected to be useful in understanding the Mott transition happening beyond two space dimensions.
\end{abstract}

\maketitle

\section{Introduction} \label{intr}
It is still a challenge for condensed matter community to describe the transition from conventional
Landau Fermi liquid to poorly understood Mott insulator in spite of intense theoretical and experimental studies due to the intricate interplay between metal-insulator transition and ubiquitous magnetic orders. Thus it is rather difficult to treat these two distinct behaviors on an equal footing. Some recent theoretical and experimental works have proposed a possible way to tackle this problem, namely, a possible quantum spin liquid Mott state is obtained at first, then the magnetic ordering is considered to be a secondary effect due to a low-temperature instability of the putative quantum spin liquids. \cite{Senthil1,Senthil2,Vojta,Custers1,Custers2} The proposed quantum spin liquids or its derivatives, the fractional fermi liquids, \cite{Senthil1,Senthil2} do not break any lattice or spin rotation symmetry and may be good candidates for exotic nonmagnetic insulator or metallic states observed in certain organic or heavy fermions compounds. \cite{Powell,Custers1,Custers2,Matsumoto} Therefore, the newly proposal provides a good way to study the nature of Mott metal-insulator transition without extra complexity from broken magnetic long-range ordering.

Generically, the Mott transition from Fermi liquid to a quantum spin liquid is described by slave-particle approaches. \cite{Wen,Florens,Lee,Senthil,Podolsky} In such an approach, the electron is firstly fractionalized into some elementary excitations like spinon, holon, and roton, and so on and then these fractional excitations are bounded with emergent gauge fields to meet the local constraints and to calculate the physical observables.

Recently, an alternative dual description of Mott transition in two space dimensions has been proposed by Mross and Senthil \cite{Mross} based on
the insight from theory of the Mott transition of bosons. In their treatment, low-lying excitations around the two dimensional Fermi surface
are first described by bosonic hydrodynamical modes, then topological defects, namely, the vortices, are identified and inserted into the effective
action of charge phase bosons which are just the hydrodynamical modes being responsible for low energy charge excitations of original Fermi liquid.
If vortices condense, the resulting Mott state is a quantum spin liquid with an emergent gapless $U(1)$ gauge field coupled to a spinons' Fermi
surface while gapped vortices (not condensate) only lead to conventional Landau Fermi liquid.

While their approach only has been applied to systems with two space dimensions, it is interesting to ask a question whether their approach can be extended to higher dimensions, specially physically interesting dimension of three. At a first sight, it seems straightforward to do it. However, a dual description in three space dimensions is rather different from its counterpart in 2D because the point-particle features of topological excitations in 2D are lost in 3D and the conventional boson-vortex duality has to be reformulated with a formidable string field theory instead of a transparent scalar-QED description (a fictitious charged superconductor with a fluctuating $U(1)$ gauge field). Since the description of the charged superconductor is crucial for the conventional boson-vortex duality, it is still unclear whether a reasonable dual description of Fermi liquid exits in the presence of topological defects whose dual description is bosonic string. In the present paper we attempt to answer these questions. We find the dual description of Fermi liquid is still robust. Different to the point-particle feature of the topological excitations in 2D, the topological defects in 3D is line singularities named vortex lines or loops. \cite{Franz,Beekman}  Therefore, after identifying the vortex line excitations in charge phase bosons, we obtain an effective description of low-lying Fermi surface fluctuations when refermionizing the corresponding bosonic hydrodynamical action in 3D. The resulting description of three dimensional Mott transition naturally incorporates both conventional Landau Fermi liquid and the expected quantum spin liquid which is formulated in terms of an effective $U(1)$ gauge field minimally coupled to spinons' Fermi surface. Moreover, a doublon metal \cite{Mross} which describes a spin gapped metal with a superconducting instability  can also be obtained when the topological defects condense in spin phase degrees of freedom. Interestingly, based on our dual approach, a new state named doubled $U(1)$ gauge theory can occur when vortex lines condense in both phase bosons. This is a consequence of two effective $U(1)$ gauge fields with decoupling of spin-opposite fermions due to their opposite gauge charges.

The remainder of this paper is organized as follows. In Sec.\ref{sec2}, we briefly review basic concepts of boson-vortex duality in four space-time dimensions for bosonic Mott transition and argue the irrelevance of complicated formulism of dual strings description in discussing the duality of three dimensional Fermi liquid in the next section in which we follow the procedure of Mross and Sethil to establish the required hydrodynamical equation for three dimensional Fermi
liquid. Its corresponding bosonized effective action for low-lying excitation of Fermi liquid is derived in Sec.\ref{sec4}. Then the topological vortex
lines excitations are identified in the three dimensional Fermi liquid and incorporated into the charge phase bosons, which leads to an effective description
of Fermi liquid and an anticipated quantum spin liquid in terms of emergent $U(1)$ gauge theory. In addition, we also find a spin gapped metal, a doublon metal, which has an instability to the superconducting pairing below a corresponding critical temperature when topological defects condense in spin phase degree of freedom.  Furthermore, when vortex lines condense in both phase bosons, an exotic doubled U(1) gauge theory emerges. Finally, a concise conclusion is devoted to Sec.\ref{sec5}.

\section{Review of boson-vortex duality in four space-time dimensions} \label{sec2}
The conventional boson-vortex duality for bosonic Mott transition in two space dimensions is rather well understood theoretically which links the classic Ginzburg-Landau-Wilson (GLW) paradigm to a fictitious charged superconductor in an emergent fluctuating $U(1)$ gauge field. \cite{Dasgupta,Fisher} The condensed state of the charged superconductor is dual to Mott insulator state of the GLW model which is characterized with the condensation of vortices in the phase degrees of freedom of the effective quantum (2+1)D XY model, a low energy limit of original GLW model. On the other hand, the disordered state of the fictitious superconductor corresponds to the superfluid state of the GLW or quantum XY model and here vortices excitations are gapped with only phonon-like gapless excitations surviving in low energy spectrum.

In the three space dimensions case, the situation is quite different. The vortices form oriented lines or loops, thus the point-particle features of vortices in two dimensional systems are lost and classic boson-vortex duality has to be modified for taking those vortex lines excitations into account. Progress has been made to solve this problem and the resulting dual description involves the celebrated Nambu-Goto, more specifically, the Nielsen-Olesen-type strings which are able to formulate the line singularity of vortices and minimally coupled to the 2-form Kalb-Ramond gauge field. \cite{Franz,Beekman} However, as we will argue here, if we only focus on the original quantum XY model and identify vortex lines in the phase description without immersing into details of duality, it is unnecessary to strive with the formidable string-gauge theory. Instead, we can use the modified quantum XY model-like theory to
proceed our discussion of dual description of three dimensional Fermi liquid-Mott insulator transition.

For our purpose, we begin with the quantum XY model in three space dimensions, which is an effective model of original GLW
action for superfluid-Mott insulator transition with integer-filling:
\begin{equation}
S=\frac{J}{2}\int d\tau d^{3}\vec{x} (\partial_{\mu}\theta(\vec{x},\tau))^{2} \label{eq1}
\end{equation}
where $J$ represents the phase stiffness, $\theta$ is the phase canonically conjugated with local particle density and $\mu = \{\vec{x},\tau\}$.

It is straightforward to use the familiar Hubbard-Stratonovich transformation to decouple the quadratic term in the original quantum XY
model, and Eq. (\ref{eq1}) becomes
\begin{equation}
S=\int d\tau d^{3}\vec{x} \left(\frac{J_{\mu}^{2}}{2J}+iJ_{\mu}\partial_{\mu}\theta\right). \label{eq2}
\end{equation}
If we decompose the phase into a smooth part $\theta_{0}$ and a singular part $\theta_{V}$ which represents the anticipated
vortex lines, we have
\begin{equation}
S=\int d\tau d^{3}\vec{x} \left(\frac{J_{\mu}^{2}}{2J}+iJ_{\mu}(\partial_{\mu}\theta_{0}+\partial_{\mu}\theta_{V})\right). \label{eq3}
\end{equation}
Here we note if one integrates over the auxiliary field $J_{\mu}$, a modified quantum XY model reads
\begin{equation}
S=\frac{J}{2}\int d\tau d^{3}\vec{x} (\partial_{\mu}\theta_{0}+a_{\mu})^{2}. \label{eq4}
\end{equation}
Since the gradient of vortex lines $\partial_{\mu}\theta_{V}$ is a singular function we have replaced it by an effective gauge field $a_{\mu}$.
A cautious reader may wonder why one introduces topological excitations like vortices or vortex lines into quantum XY model above and what role
those topological objects play. The reason is that in the original XY action, the phase variables do not remember they are phase defined mod $2\pi$,
which means the canonically conjugated local density of particles cannot be quantizied into integer per area or volume. Thus, the phase-only description
of original quantum XY action is unable to describe the Mott state. Therefore, in order to have a description of Mott insulator,
one has to impose topological defects such as vortices or vortex lines into the XY model due to their ability to implement the constraint of mod $2\pi$ for
phase variables. Only those topological defects condense, the quantization of density of particles will be fulfilled, which leads to an expected Mott insulator.

Obviously, this modified quantum XY model is general for arbitrary space dimensions because we do not refer to any specific realization
of duality but only make a substitution for incorporating topological vortex lines excitations. Thus, in the discussion of dual approach for
Fermi liquid-Mott insulator transition in the remaining parts of the present paper, it could be safe to follow the same logic as it has been done in this
section without referring specific realization of duality.

For self-contained goal, we would like to state the main results of boson-vortex duality in four space-time dimensions for the condensed matter
community, but eager readers may skip this part without loss of understanding of other sections in this paper.
We continue with the quantum XY action including the auxiliary field $J_{\mu}$, and one can integrate over the smooth phase field $\theta_{0}$, which
leads to a constraint $\partial_{\mu}J_{\mu}=0$. It has been emphasized by Franz \cite{Franz} in three space dimensions that this constraint can only be fulfilled with $J_{\mu}=\epsilon_{\mu\nu\alpha\beta}\partial_{\nu}B_{\alpha\beta}$
where $B_{\mu\nu}$ is a 2-form antisymmetric gauge field and $\epsilon_{\mu\nu\alpha\beta}$ a totally antisymmetric tensor. Therefore, one gets
\begin{equation}
S=\int d\tau d^{3}\vec{x} \left(\frac{H_{\alpha\beta\gamma}^{2}}{3J}+iB_{\alpha\beta}\epsilon_{\mu\nu\alpha\beta}\partial_{\mu}\partial_{\nu}\theta_{V}\right)
\end{equation}
where $H_{\alpha\beta\gamma}=\partial_{\alpha} B_{\beta\gamma}+\partial_{\beta} B_{\gamma\alpha}+\partial_{\gamma} B_{\alpha\beta}$ is a field-strength tensor.
To proceed the dual transformation, the vortex lines degrees of freedom are replaced by the Nielsen-Olsen-type strings \cite{Zwiebach} and their effective action is
\begin{eqnarray}
&& S=\int d\tau d^{3}\vec{x} \frac{H_{\alpha\beta\gamma}^{2}}{3J} + \int \emph{D}[X]\int d\sigma\sqrt{h} \nonumber\\
&&\hspace{0.5cm} \left(\left|\left(\frac{\delta}{\delta\Sigma_{\mu\nu}}
-2\pi B_{\mu\nu}\right)\Phi[X]\right|^{2}+M_{eff}^{2}|\Phi[X]|^{2}\right) \nonumber\\
&& \hspace{0.5cm}+S'_{int},
\end{eqnarray}
where $X$ is the so-called string coordinate, $\sigma$ is the string parameter, $\Phi$ is the string field, $\Sigma_{\mu\nu}$ is the surface element of a worldsheet and $h$ can be considered as an induced metric of the strings
in three space dimensions. Here a mass term $M_{eff}$ and a remaining interacting term $S'_{int}$ which represents short-range string interaction are added.
The physical picture is clear in this situation in spite of the formidable formulism of string-gauge action above. \cite{Franz} The dual description involves the Nielsen-Olesen strings with intrinsic thickness defined by the vortices core sizes and a 2-form
effective gauge field named Kalb-Ramond gauge field in string theory. Therefore, the disorder phase of original GLW or quantum XY action is described in terms of condensation of strings while the original superfluid state corresponds to gapped state of those Nielsen-Olesen strings with free Kalb-Ramond gauge field supporting gapless photonic excitations which in fact play a role as the Goldstone modes
in original symmetry breaking phase of GLW or XY models.

\section{Hydrodynamical description of Fermi liquid} \label{sec3}
Having reviewed the basic of boson-vortex duality, we begin our discussion of the dual description of Fermi liquid-Mott insulator
transition in three space dimensions. Following Mross and Senthil, \cite{Mross} we start with the hydrodynamical equation for conventional Fermi liquid in 3D as follows
\begin{eqnarray}
&& \left(\frac{\partial}{\partial t}+\vec{v}_{\vec{k}}\cdot\frac{\partial}{\partial \vec{x}}\right)\delta n_{\vec{k}\sigma}(\vec{x},t) \nonumber\\
&& \hspace{0.5cm}+\frac{1}{L^{d}}\sum_{\vec{k}'\sigma'}f_{\vec{k}\sigma,\vec{k}'\sigma'}\delta(k_{F}-|\vec{k}|)\hat{\vec{k}}\cdot \frac{\partial}{\partial \vec{x}}\delta n_{\vec{k}'\sigma'}(\vec{x},t)=0,
\end{eqnarray}
where $k_{F}$ is Fermi momentum, $\vec{v}$ is the velocity of quasi-particle with momentum $\vec{k}$, $\delta n_{\vec{k}\sigma}(\vec{x},t)$ is the deviation of ground state distribution function and $f_{\vec{k}\sigma,\vec{k}'\sigma'}$ is the familiar Landau interaction function. Then we introduce local density and total density function in 3D
\begin{equation}
\rho_{\sigma}(\Omega,\vec{x},t)\equiv\int_{-\Lambda}^{+\Lambda}\frac{dk}{(2\pi)^{3}}\delta n_{k,\Omega,\sigma}(\vec{x},t),
\end{equation}
where $\Lambda$ is a cutoff momentum. Here $\delta n_{\vec{k},\sigma}(\vec{x},t)=\delta n_{k,\Omega,\sigma}(\vec{x},t)$, $\Omega\equiv(\theta,\varphi)$,  $d\Omega\equiv\int_{0}^{2\pi}d\varphi\int_{0}^{\pi}d\theta\sin\theta$, and
\begin{equation}
\rho_{\sigma}(\vec{x},t)\equiv\int d\Omega k_{F}^{2}(\Omega)\rho_{\sigma}(\Omega,\vec{x},t).
\end{equation}
Consequently, multiplying above hydrodynamical equation with $1/(2\pi)^{3}$ and integrating over $k$, one obtains
\begin{eqnarray}
&& \left(\frac{\partial}{\partial t}+\vec{v}_{F}(\Omega)\cdot\frac{\partial}{\partial \vec{x}}\right)\rho_{\sigma}(\Omega,\vec{x},t) + \frac{1}{(2\pi)^{3}}\sum_{\sigma'}\int d\Omega' k_{F}^{2}(\Omega')\nonumber\\
&& \hspace{1cm}\times f_{\sigma\sigma'}(\Omega,\Omega')\hat{\vec{k}}_{F}(\Omega)\cdot\frac{\partial}{\partial \vec{x}}\rho_{\sigma'}(\Omega',\vec{x},t)=0.
\end{eqnarray}
If we use the notation $\vec{v}_{F}\cdot\frac{\partial}{\partial \vec{x}}=v_{F}\hat{\vec{v}}_{F}\cdot\frac{\partial}{\partial \vec{x}}\equiv v_{F}\partial_{\parallel}$ with $\partial_{\parallel}\equiv\hat{\vec{v}}_{F}(\Omega)\cdot\frac{\partial}{\partial \vec{x}}$ ($\hat{\vec{v}}_{F}$ or $\hat{\vec{k}}_{F}$ is unit vector), we can further simplify above expression into the following form
\begin{eqnarray}
&&\left(\frac{\partial}{\partial t}+v_{F}\partial_{\parallel}\right)\rho_{\sigma}(\Omega,\vec{x},t)\nonumber\\
&& +\frac{1}{(2\pi)^{3}}\sum_{\sigma'}\int d\Omega' k_{F}^{2}(\Omega')f_{\sigma\sigma'}(\Omega,\Omega')\partial_{\parallel} \rho_{\sigma'}(\Omega',\vec{x},t)=0.
\end{eqnarray}
Now we turn into a phase bosons description of Fermi liquid with $\rho_{\sigma}(\Omega,\vec{x},t)=\frac{1}{2\pi}\partial_{\parallel}\phi_{\sigma}(\Omega,\vec{x},t)$, so the equation of motion for Fermi liquid is transformed into
\begin{eqnarray}
&&\left(\frac{\partial}{\partial t}+v_{F}\partial_{\parallel}\right)\partial_{\parallel}\phi_{\sigma}(\Omega,\vec{x},t)\nonumber\\
&& +\frac{1}{(2\pi)^{3}}\sum_{\sigma'}\int d\Omega' k_{F}^{2}(\Omega')f_{\sigma\sigma'}(\Omega,\Omega')\partial_{\parallel}\partial_{\parallel}'\phi_{\sigma}(\Omega',\vec{x},t) = 0.
\end{eqnarray}

\section{The effective bosonized action for Fermi liquid} \label{sec4}
In this section, we proceed to discuss an effective bosonization action for Fermi liquid in 3D which is appropriate for dual description of Mott transition.
It is easy to see that the following action can give rise to the hydrodynamical equation for Fermi liquid in 3D derived in the last section
\begin{equation}
S=\int dt\int d^{3}\vec{x}L(\partial_{t}\phi_{\sigma},\partial_{\parallel}\phi_{\sigma}), \;\; L=L_{0}+L_{I},
\end{equation}
where $L_{0}=\sum_{\sigma}\int d\Omega k^{2}_{F}(\Omega) [\partial_{t}\phi_{\sigma}(\Omega,\vec{x},t) \partial_{\parallel}\phi_{\sigma}(\Omega,\vec{x},t) - v_{F}(\partial_{\parallel}\phi_{\sigma}(\Omega,\vec{x},t))^{2}]$ and $L_{I} = -\frac{1}{(2\pi)^{3}} \sum_{\sigma\sigma'}  \int d\Omega k^{2}_{F}(\Omega)$ $\times \int d\Omega' k^{2}_{F}(\Omega') f_{\sigma\sigma'}(\Omega,\Omega') \partial_{\parallel}\phi_{\sigma}(\Omega,\vec{x},t) \partial'_{\parallel}\phi_{\sigma'}(\Omega',\vec{x},t)$.
Consider $\phi_{\sigma}(\Omega,\vec{x},t)$ as the canonical coordinate, its corresponding canonical momentum can be written as $\Pi_{\sigma}(\Omega,\vec{x},t)\equiv\frac{\partial L}{\partial(\partial_{t}\phi_{\sigma}(\Omega,\vec{x},t))}=\sin\theta k^{2}_{F}(\Omega)\partial_{\parallel}\phi_{\sigma}(\Omega,\vec{x},t)$,
therefore we may identify a commutation relation for these two canonically conjugated variables when we turn to the operator framework
\begin{equation}
[\hat{\phi}_{\sigma}(\Omega,\vec{x},t),\hat{\Pi}_{\sigma}(\Omega',\vec{x}',t)]=i\delta(\theta-\theta')\delta(\varphi-\varphi')\delta(\vec{x}-\vec{x}')
\end{equation}
or we have
\begin{eqnarray}
&& [\hat{\phi}_{\sigma}(\Omega,\vec{x},t),2\pi\sin\theta' k^{2}_{F}(\Omega')\hat{\rho}_{\sigma}(\Omega',x,t)] \nonumber\\
&& \hspace{1cm} =i\delta(\theta-\theta')\delta(\varphi-\varphi')\delta(\vec{x}-\vec{x}').
\end{eqnarray}
It is straightforward to derive a further commutation relation if one defines the local particle density operator ($\hat{\rho}_{\sigma}(\vec{x},t)\equiv\int d\Omega k_{F}^{2}(\Omega)\hat{\rho}_{\sigma}(\Omega,\vec{x},t)$)
 and the local phase operator ($\hat{\phi}_{\sigma}(\vec{x},t)=\frac{1}{4\pi}\int d\Omega \hat{\phi}_{\sigma}(\Omega,\vec{x},t)$)
\begin{equation}
[\hat{\phi}_{\sigma}(\vec{x},t),2\pi \hat{\rho}_{\sigma}(\vec{x}',t)]=i\delta(\vec{x}-\vec{x}').
\end{equation}
This commutation relation will be useful in next section where we introduce topological objects named vortex lines.
Having the real-time action in hand, we can also transform it into imaginary-time action which reads as follows
\begin{equation}
S=\int d\tau\int d^{3}\vec{x}L(\partial_{\tau}\phi_{\sigma},\partial_{\parallel}\phi_{\sigma}), \;\; L=L_{0}+L_{I}
\end{equation}
where $ L_{0}=\sum_{\sigma}\int d\Omega k^{2}_{F}(\Omega)
[-i\partial_{\tau}\phi_{\sigma}(\Omega,\vec{x},\tau)\partial_{\parallel}\phi_{\sigma}(\Omega,\vec{x},\tau)$ $+ v_{F}(\partial_{\parallel}\phi_{\sigma}(\Omega,\vec{x},\tau))^{2}]$ and $ L_{I}=\frac{1}{(2\pi)^{3}}\sum_{\sigma\sigma'}\int d\Omega\int d\Omega' $ $k^{2}_{F}(\Omega) k^{2}_{F}(\Omega')f_{\sigma\sigma'}(\Omega,\Omega')\partial_{\parallel}\phi_{\sigma}(\Omega,\vec{x},\tau)\partial'_{\parallel}\phi_{\sigma'}(\Omega',\vec{x},\tau)$.
As done in the situation of the hydrodynamical description for one dimensional fermions systems (i.e. the Luttinger liquid), \cite{Gogolin} one can define charge and spin phase variables (bosons)
\begin{eqnarray}
&&\phi_{c}(\Omega,\vec{x},\tau)=\frac{1}{2}(\phi_{\uparrow}(\Omega,\vec{x},\tau)+\phi_{\downarrow}(\Omega,\vec{x},\tau)), \\
&&\phi_{s}(\Omega,\vec{x},\tau)=\frac{1}{2}(\phi_{\uparrow}(\Omega,\vec{x},\tau)-\phi_{\downarrow}(\Omega,\vec{x},\tau)).
\end{eqnarray}
Correspondingly, one can also introduce charge and spin density variables
\begin{eqnarray}
&&\rho_{c}(\Omega,\vec{x},\tau)=\frac{1}{2}(\rho_{\uparrow}(\Omega,\vec{x},\tau)+\rho_{\downarrow}(\Omega,\vec{x},\tau)), \\
&&\rho_{s}(\Omega,\vec{x},\tau)=\frac{1}{2}(\rho_{\uparrow}(\Omega,\vec{x},\tau)-\rho_{\downarrow}(\Omega,\vec{x},\tau)),
\end{eqnarray}
where $\rho_{c}=\frac{1}{2\pi}\partial_{\parallel}\phi_{c}$ and $\rho_{s}=\frac{1}{2\pi}\partial_{\parallel}\phi_{s}$.
Using above charge and spin phase variables, the effective action for Fermi liquid in 3D can be regrouped into three terms which represent charge, spin excitations, and their interactions, respectively,
\begin{eqnarray}
&& S=S_{c}+S_{s}+S_{c-s}, \\
&& S_{c}=\int d\tau\int d^{3}\vec{x}(L_{c}^{0}+L_{c}^{I}), \\
&& S_{s}=\int d\tau\int d^{3}\vec{x}(L_{s}^{0}+L_{s}^{I}), \\
&& S_{c-s}=\frac{1}{4\pi^{3}}\int d\Omega\int d\Omega' k^{2}_{F}(\Omega) k^{2}_{F}(\Omega')\nonumber\\
&& \hspace{1cm} \times (f_{\uparrow\uparrow}-f_{\downarrow\downarrow})\partial_{\parallel}\phi_{c}(\Omega,\vec{x},\tau)\partial'_{\parallel}\phi_{s}(\Omega',\vec{x},\tau).
\end{eqnarray}
Here the notations $L_{c,s}^{0}, L_{c,s}^{I}$ are introduced as
$ L_{c}^{0}=\int d\Omega k^{2}_{F}(\Omega) [-i\partial_{\tau}\phi_{c}(\Omega,\vec{x},\tau)\partial_{\parallel}\phi_{c}(\Omega,\vec{x},\tau)+v_{F}(\partial_{\parallel}\phi_{c}(\Omega,\vec{x},\tau))^{2}]$, $L_{c}^{I}=\frac{1}{4\pi^{3}}\int d\Omega\int d\Omega' k^{2}_{F}(\Omega) k^{2}_{F}(\Omega')$ $f_{s}(\Omega,\Omega') \partial_{\parallel}\phi_{c}(\Omega,\vec{x},\tau)\partial'_{\parallel}\phi_{c}(\Omega',\vec{x},\tau)$,  $L_{s}^{0}=\int d\Omega k^{2}_{F}(\Omega)
$ $[-i\partial_{\tau}\phi_{s}(\Omega,\vec{x},\tau)\partial_{\parallel}\phi_{s}(\Omega,\vec{x},\tau)+v_{F}(\partial_{\parallel}\phi_{s}(\Omega,\vec{x},\tau))^{2}]$, $L_{s}^{I}=\frac{1}{4\pi^{3}}\int d\Omega\int d\Omega' k^{2}_{F}(\Omega) k^{2}_{F}(\Omega')f_{a}(\Omega,\Omega') \partial_{\parallel}\phi_{s}(\Omega,\vec{x},\tau) $ $\partial'_{\parallel}\phi_{s}(\Omega',\vec{x},\tau)$,
and $f_{s}=(f_{\uparrow\uparrow}+f_{\downarrow\uparrow})/2$ and $f_{a}=(f_{\uparrow\uparrow}-f_{\uparrow\downarrow})/2$ are standard Landau interaction function, respectively. In particular, if we further assume a paramagnetic solution of Fermi liquid, the charge-spin interaction term will vanish in this situation and we will obtain an effective action with seemingly spin-charge separation formulism.
In the remaining part of the present paper, we only consider this paramagnetic Fermi liquid for simplicity.

\section{Incorporating vortex lines into phase bosons} \label{sec5}
It is well-known that vortices are favorable in classical or quantum XY model-like systems in space dimension of two when thermal or
quantum fluctuation are strong enough to suppress ordered states. If these vortices proliferate, they will drive the whole system into thermal
or quantum disordered phase and long-range order will disappear. Particularly, if the original bosonic Hamiltonian filled with integral number of particles,
the vortices condensation induced transition should be superfluid-Mott transition with Mott state identified by condensate of vortices. However,
in 3D point-particle-like vortices are not energetically preferable while topological vortex lines excitation may drive the
Mott transition of bosons. We have briefly reviewed the basis of the dual approach of this Mott transition in the previous section and based on those
insights we will follow the same logic of Mross and Senthil \cite{Mross} to fulfill
our dual description of Fermi liquid-Mott insulator transition in space dimension of three in this section.

As mentioned above, Fermi liquid can be rewritten in terms of bosonized effective action which only
incorporates low energy charge and spin excitation around Fermi surface. Based on insights from the above discussion of vortex lines driven superfluid-Mott transition of bosons and stimulated by work of Mross and Senthil for two dimensional Fermi liquid-Mott insulator transition, we also
expect that a similar Fermi liquid-Mott insulator transition could be driven by condensation of vortex lines in space dimension of three. Now we first proceed to identify vortex lines in charge phase bosons $\phi_{c}$. As emphasized by Mross and Senthil, \cite{Mross} the correct procedure of incorporating topological object is to firstly identify local charge phase which is canonically conjugated to local charge density, then to insert topological objects such as vortices or vortex lines into local charge phase. It is easy to see the local charge phase is a simple addition of local phase for different spins
\begin{equation}
\phi_{c}(\vec{x},\tau)=\frac{1}{2}(\phi_{\uparrow}(\vec{x},\tau)+\phi_{\downarrow}(\vec{x},\tau))=\frac{1}{4\pi}\int d\Omega \phi_{c}(\Omega,\vec{x},t).
\end{equation}
Here it is useful to expand $\phi_{c}(\Omega,\vec{x},t)$ in terms of spherical harmonic functions $Y_{lm}(\Omega)$
\begin{equation}
\phi_{c}(\Omega)=\sum_{lm}Y_{lm}(\Omega)\phi_{c,lm},
\end{equation}
where $\phi_{c,lm}=\frac{1}{4\pi}\int d\Omega Y^{\ast}_{lm}(\Omega)\phi_{c}(\Omega)$. We note that $\phi_{c,l=m=0}=\frac{1}{4\pi}\int d\Omega \phi_{c}(\Omega)$ is just the local charge phase we need. Therefore, following Mross and Senthil, we need to separate the mode with $l=m=0$ from other modes in the bosonic effective action. After a straightforward manipulation, we obtain the following action
\begin{equation}
S_{c}=\int d\tau\int d^{3}\vec{x}L_{c},
\end{equation}
where
\begin{widetext}
\begin{equation}
L_{c}=-i\sum_{lml'm'\neq0}\partial_{\tau}\phi_{c,lm}\vec{Z}_{mm'}^{ll'}\partial_{\vec{x}}\phi_{c,l'm'}-2i\sum_{lm\neq0}\partial_{\tau}\phi_{c,00}\vec{Z}_{m0}^{l0}\partial_{\vec{x}}\phi_{c,lm}
+\sum_{lm,l'm'}\partial_{\alpha}\phi_{c,lm} M_{lm,l'm'}^{\alpha\beta}\partial_{\beta}\phi_{c,l'm'}.
\end{equation}
Here $\vec{Z}_{mm'}^{ll'}=\int d\Omega k_{F}^{2}(\Omega)Y_{lm}(\Omega)Y_{l'm'}(\Omega)\hat{\vec{v}}_{F}(\Omega)$ and $M_{lm,l'm'}^{\alpha\beta}=2v_{F}\int d\Omega k_{F}^{2}(\Omega)\hat{v}_{F}^{\alpha}(\Omega)\hat{v}_{F}^{\beta}(\Omega)Y_{lm}(\Omega)Y_{l'm'}(\Omega)+\frac{1}{4\pi^{3}}\int d\Omega\int d\Omega'k_{F}^{2}(\Omega)k_{F}^{2}(\Omega')f_{s}(\Omega,\Omega')\hat{v}_{F}^{\alpha}(\Omega)\hat{v}_{F}^{\beta}(\Omega')Y_{lm}(\Omega)Y_{l'm'}(\Omega')$.
Consequently, we insert the vortex lines degrees of freedom into the mode of $l=m=0$ with the direct substitution $\phi_{c,lm}\rightarrow\phi_{c,lm}+\phi_{V}\delta_{l=m=0}$. Since the vortex lines are singular manifolds, we replace its gradient $\partial_{\mu}\phi_{V}$ with an effective $U(1)$ gauge field $a_{\mu}$.
Therefore an effective action for charge degrees of freedom of Fermi liquid can be derived as
\end{widetext}
\begin{equation}
S_{c}=\int d\tau\int d^{3}\vec{x} \int d\Omega k_{F}^{2} (L\{\partial_{\tau}\phi_{c},\partial_{\parallel}\phi_{c}\}+j_{\mu}a_{\mu})
\end{equation}
and for spin section, the effective action can be written as
\begin{equation}
S_{s}=\int d\tau\int d^{3}\vec{x} \int d\Omega k_{F}^{2} (L\{\partial_{\tau}\phi_{s},\partial_{\parallel}\phi_{s}\}).
\end{equation}
Combining above two parts for the whole effective action of Fermi liquid, we have
\begin{equation}
S=\int d\tau\int d^{3}\vec{x} \int d\Omega k_{F}^{2} (L\{\partial_{\tau}\phi_{\sigma},\partial_{\parallel}\phi_{\sigma}\}+ j_{\mu}^{\sigma}a_{\mu}).
\end{equation}
It is well know that the action without vortex lines describes conventional Fermi liquid if one refermionizes the bosonic
action by using bosonization identity $f_{\sigma}(\vec{x},\tau,\Omega)\sim e^{i(2\pi)^{3/2}\phi_{\sigma}(\vec{x},\tau,\Omega)}$ with $f_{\sigma}$ representing a fermion with a spin. As a matter of fact, we find a more general bosonization relation for refermionizing phase bosons of original Fermi liquid situated in arbitrary dimension $d$ with its form being $f_{\sigma}(\vec{x},\tau,\Omega)\sim e^{i(2\pi)^{d/2}\phi_{\sigma}(\vec{x},\tau,\Omega)}$. In the low energy limit, all interaction terms in bosonic action between bosons of two patches near Fermi surface is irrelevant and
we may use free fermions' Lagrangian for the term without vortex lines
\begin{equation}
L\{\partial_{\tau}\phi_{\sigma},\partial_{\parallel}\phi_{\sigma}\}=f_{\sigma}^{+}\left(\partial_{\tau}-\mu_{F}+\frac{-\nabla^{2}}{2m}\right)f_{\sigma}
\end{equation}
and the whole Lagrangian can be rewritten as
\begin{equation}
L=f_{\sigma}^{+}\left(\partial_{\tau}-\mu_{F}-ia_{0}+\frac{-(\nabla-i\vec{a})^{2}}{2m}\right)f_{\sigma},
\end{equation}
where $\mu_{F}$ and $m$ are effective chemical potential and mass of the fermions, respectively. We notice that the above action is just the one obtained in continuous Mott transition for generic systems described by Hubbard model
nearly half-filling \cite{Podolsky} and Kondo-breakdown mechanism \cite{Senthil2,Vojta,Paul,Pepin} for heavy fermions compounds like $YbRh_{2}Si_{2}$ and $YbRh_{2}Si_{2-x}Ge_{x}$ \cite{Custers1,Custers2} in 3D with the help of slave rotor or slave boson representations. This action describes fermion interacting with each other in terms of an effective $U(1)$ gauge field
in the background of a Fermi surface of those fermions. In fact, these fermions are fermionic spinons without carrying charges due to
condensation of vortex lines in charge phase bosons. The condensation gaps the charge excitation in low energy and the whole theory describe
a $U(1)$ quantum spin liquid with a spinons' Fermi surface. This picture has been proposed to describe the nonmagnetic state down to lowest temperature
in organic Mott insulator. \cite{Lee,Powell}

It should be emphasized that the above $U(1)$ gauge theory action implicitly assumes a deconfined state for the $U(1)$ gauge field which is believed to be possible in 3D. This is in contrast to the situation of 2D where de-confinement is only faithfully established near
some bosonic quantum critical point besides various artificial large-N arguments in the presence of Fermi surface. \cite{Kogut,Polyakov1,Polyakov2,Senthil3,Senthil4,Kim,Lee2} Otherwise, if $U(1)$ gauge field is confined,
the above action is actually meaningless because only bound states of fermions are able to appear in low energy spectrum due to the linear confined potential mediated by confined gauge field between a pair of fermions.
Therefore, the insulating state with confined gauge field cannot be described by a quantum spin liquid but may be a valence bond solid or more conventional spin ordered phases like collinear anti-ferromagnetic states. \cite{Sachdev1,Sachdev2}

Moreover, a doublon metal obtained by Mross and Senthil \cite{Mross} with a spin gap can also be found when vortex lines condense in spin but not in charge phase bosons in three space dimensions. This metal state is described by the Lagrangian
\begin{equation}
L=\sum_{\sigma=\pm1} f_{\sigma}^{+}\left(\partial_{\tau}-\mu_{F}-i\sigma a_{0}+\frac{-(\nabla-i\sigma\vec{a})^{2}}{2m}\right)f_{\sigma},
\end{equation}
where fermions $f_{\sigma}$ with distinct spins couple to a gauge field $b_{\mu}$ with opposite gauge charges, thus gives rise to an attractive interaction between opposite charged fermions. The doublon metal mentioned above is originally derived from a fluctuating doped anti-ferromagnet, which has been applied to the cuprate superconductors. \cite{Kaul,Galitski,Moon} Although a paring instability will be induced by the mediated attractive interaction in this model, we could consider this metal state as a nontrivial intermediate temperature state above the superconducting critical temperature, which is a non-Fermi liquid state with anomalous thermal and transport behaviors compared to conventional Landau Fermi liquid.

Up to now, we have found that three distinct states by our dual approach, they are the conventional Fermi liquid, a $U(1)$ quantum spin liquid and a doublon metal, respectively. Since vortex lines could condense in charge or spin phase bosons, it may be reasonable to expect a double condensation in both phase bosons. It is straightforward to find a doubled gauge theory, which reads as follows:
\begin{eqnarray}
&& L=f_{\uparrow}^{+}\left(\partial_{\tau}-\mu_{F}-i(a_{0}+b_{0})+\frac{-(\nabla-i(\vec{a}+\vec{b}))^{2}}{2m}\right)f_{\uparrow} \nonumber \\
&& +f_{\downarrow}^{+}\left(\partial_{\tau}-\mu_{F}-i(a_{0}-b_{0})+\frac{-(\nabla-i(\vec{a}-\vec{b}))^{2}}{2m}\right)f_{\downarrow}.
\end{eqnarray}
In this Lagrangian, fermions with different spins have the same gauge charges when couple to gauge field $a_{\mu}$ but carry opposite gauge charges with respect to gauge field $b_{\mu}$. Since the obtained Lagrangian is a new result of this paper, here we would like to analyze its physical properties briefly. Firstly, at mean field level where one simply neglects those two gauge fields, an expected gas of free fermions is reproduced. Then, when one turns up the coupling between fermions and two kinds of gauge fields, interestingly, the effective interaction vanishes between fermions with opposite spins mediated by gauge field $a_{\mu}$ and $b_{\mu}$, due to difference in the sign of gauge charges of fermions at the one loop level. Thus, fermions with opposite spins are approximately decoupled to the leading order of perturbation theory and the remaining parts of the effective Lagrangian can be treated separately as done in the Kondo-breakdown mechanism \cite{Senthil2,Vojta,Paul,Pepin} of heavy fermions systems. To our knowledge, this effective action has not been realized by any realistic lattice models, but we hope this exotic quantum state may be useful for further study.

It is noted recently, the derivative of strings theory, the anti-de Sitter/conformal field theory (AdS/CFT) correspondence, has been used to attack the perplexing non-Fermi liquid behaviors in strongly correlated electronic systems \cite{Faulkner1,Faulkner2,Liu} and a clear correspondence is found between certain quantum gravity theories with an emergent AdS$_{2}\times R^{d-1}$ geometry and fractionalized metallic states of Kondo-Heisenberg model, \cite{Sachdev3} thus we hope similar physics as established in terms of our dual approach above may be inspected through this promising approach.
\section{Conclusion}
In conclusion, we have extended the dual approach of two dimensional Fermi liquid-Mott insulator transition
into the physically interesting space dimension of three in spite of complicated details of the dual description with
vortex lines involved. A quantum spin liquid state has been found in three space dimensions when vortex lines condense
in charge phase bosons while the gapped vortex lines only lead to conventional Landau Fermi liquid. In addition, we also have found a spin gapped metal state called as doublon metal, which may support an unconventional superconductivity when topological defects condense in spin phase degrees of freedom.  In addition, a double condensation of vortex lines in both phase bosons has been examined and an exotic doubled $U(1)$ gauge theory emerges in this case. This result describes a decoupling of spin-opposite fermions due to their opposite gauge charges. Therefore, four distinct states have been found in our dual approach although details of these states and their corresponding subtle quantum phase transitions have not been captured by this method, which deserves further study in the future. We hope this alternative dual description may provide more insights into Fermi liquid-Mott insulator transition in space dimension of three.  Furthermore, similar physics as established in terms of our dual approach above could be inspected through the promising AdS/CFT correspondence since the holographic realization of the familiar compressible quantum states of condensed matter physics, superfluid and Fermi liquid, have been found firmly. \cite{Sachdev4}

\begin{acknowledgments}
The authors would like to thank Y. Q. Wang for useful discussion on Nambu-Goto strings and related issues of duality. The work is supported partly by the Program for NCET, NSF and the Fundamental Research Funds for the Central Universities of China.
\end{acknowledgments}

\end{document}